\documentclass[fleqn,twoside]{article}
\usepackage{gc,epsf}

\heads{V.~Folomeev, V.~Gurovich, I.~Tokareva}
      {Geometric Model of Quintessence}

\begin{document}

\twocolumn[

\Arthead{12}{2006}{2-3 (46-47)}{163}{169}

\Title{Geometric Model of Quintessence}

 \Authors{V.~Folomeev, V.~Gurovich \foom 1 }
           {and I. Tokareva \foom 2 }
           {Physics Institute of NAN KR, 265 a, Chui str., Bishkek, 720071,  Kyrgyz Republic}
           {Physics Department, Technion, Technion-city, Haifa 32000, Israel}

\Abstract

    {On basis of modification of Einstein's gravitational equations by
    adding the term $f(R)\propto \beta R^n$, a geometric model of quintessence
    is proposed. The evolution equation for the scale factor $a$ of the Universe
    is analyzed for the two parameters $n=2$ and $n=4/3$, which were preferred
    by previous studies of the early Universe. Another choice of parameters
    $n$ and $\beta$ is proposed from the following reasons:  the exponent $n$
    close to $1.2953$ follows from the request for the evolution of the Universe
    after recombination to be close to the evolution of the flat FRW model with cold
    dark matter and reasonable age of the Universe defines the value of the
    coefficient $\beta$. Such a model corresponding to the evolution of the Universe
    with the dynamical $\Lambda$-term describes well enough the observational data.}

\bigskip
]

\email 1 {astra@freenet.kg}

\email 2 {iya@tx.technion.ac.il}

\section{Introduction}
The discovery of accelerated expansion of the
 Universe \cite{Riess,Perlmutter,Tegmark} has stimulated the quest
  for mechanism of present inflation.
  The most famous theoretical model of  dark energy (DE) is the
cosmological constant $\Lambda$. The corresponding FRW
 solution for flat Universe with the present densities ratio for
  cold matter and dark energy ($\Omega_{m}/\Omega_{\Lambda0}
  \sim 0.3/0.7$) describes satisfactorily the evolution of the Universe
   at
low redshifts~\cite{Tegmark,Linde,sahni1}. However,  the nature of
the
 constant $\Lambda$-term has been remaining to be inexplicable
 during many years.

It is well-known,  the application of the constant $\Lambda$-term
for  modeling of the early Universe was initially confronted with
principal difficulties. Solving  the prob\-lems of  the very early
Universe, this term has to be reducing by several order of magnitude
during following  evolution of the Universe. This problem was solved
by rejection of $\Lambda$-term,
  and corresponding inflationary behavior was determined by models
  of ``effective $\Lambda$-term'' - quasi-classical scalar fields
  in the one way.  The progress of these
   models is well-known.
   Another way to describe the inflationary behavior
   is to take into account the polarization of
 vacuum of quantum fields in the  early Universe. Taking into
 account of the effects of  polarization leads to
 appearance of the  terms non-linear on curvature in the
 Einstein-Hilbert
 action. In such models  the inflation appears
 self-consistently. Let us note two issues:\\
- a correction to the Einstein-Hilbert  action  with an arbitrary
function of scalar curvature $R$
  is equivalent  mathematically to the
  introduction of scalar field into the classical Friedmann
  cosmology \cite{sahni1,Riess1,Barrow,Baib};\\
  - the terms of the form $R^n$ were investigated in
   the early works done on the problem of singularity before
  obtaining of exact corrections to the Einstein action follo\-wing
  from the one loop approximation
  \cite{Gur2}. The part
  of such solutions approaches asymptotically to the Friedmann solutions
  with $\Lambda$-term, however physical results of the solutions
  have   not been explained at that time.

   For the purpose to explain
   the  accelerated expansion of  the Universe today, it is
   naturally to use
   the experience
  cumulative at investigations  of the early Universe. Thus,  one of
  the tendencies is concerned with the hypothesis of existence
  some scalar fields that determine the density of dark energy
   (see e.g. \cite{Ratra,Zlatev,Starobinsky}). The another  tendency models is an effective
  quasi-hydrodynamical energy-momentum ten\-sor
  describing the observational data \cite{kamen}. And the third tendency
   consists in
  generalization of the Ein\-stein-Hilbert equation by inclusion of
   curvature inva\-riants
  \cite{Turner,Capoz,fol} analogously to  the earlier works
  men\-tioned above.
  The last approach one can consider  to be
   either  an  independent approach to describing of DE or  an
 analogue of inclusion of scalar fields (in case of $f(R)$)
  as stated above.

 In the works  on the higher order gravity theories (HOGT),
the models with power corrections were inves\-tigated, however they
have never been fitted to whole set of the observational data.

  In this paper, the model with correction $f(R)\propto R^n$ with
  $n>0$ is considered in detail for the purpose to correlate it with
 the observational data. In the other words, we would like to
 obtain the  model  that does not  conflict with the  scenario of
 the  large scale  structure formation (in past) and describes
 satisfactorily the Universe undergoing an accelerated expansion
 at present.  Therefore, at the minimum,
  within the framework of the $f(R)$-theories,
  we will obtain
 solutions  remind $\Lambda$CDM model  describing well enough by
 set of the observational data \cite{Tegmark,Linde}.
  However, as it was mentioned by various authors (see
 \cite{sahni1} and references therein), the observational data indicates
 the models of dynamical DE. Hence, our second aim is  to search out
 such dynamical solutions within the framework of
 HOGT and to find out whether  these solutions are
 preferable.

 This paper is organized as follows. In  section 2 the basic
 equations of  HOGT are presented. Section 3 is devoted
 to consideration of the models with $n=2$ and $n=4/3$. In section 4 for the
 corrections of the form $f(R)\propto \beta R^n$ we find the exponent
 $n_1=1.2953$
  which allows generalization of the Einstein equation for the scale factor
  $a$ to have a particular   solution corresponding to the flat
  FRW solution for cold  matter.
  We show that instability of the
  solutions, that are close to this particular solution at $z\gg 1$,
   may lead to  the accelerated behavior  of
  the model at present and the following asymptotic
    approach of the solution to
  the solution with the constant $\Lambda$-term.
   In section 5 we discuss our results and compare them  to
 observations. In the model there are only two free parameters - the coefficient $\beta$ and a slight deviation of the parameter
  $n$  from $n_1=1.2953$ mentioned above.
   Fixed from one set of the observational
  data, they  allow to obtain the rest of the set
  of the observational data.

\section{Basic equations}
\label{basic}
As suggested by observations, we consider the flat cos\-molo\-gical
Friedmann model with the metric
\begin{equation}
ds^2=d\tau^2-a(\tau)^2 (dx^2+dy^2+dz^2).
\end{equation}
If $H_0$ denotes the presently observable Hubble constant
 (the
subscript 0 will always indicate the present epoch), the reduced
curvature tensor $\rho_{k}^{l}\equiv H_0^{-2}R_{k}^{l}$ has the
following matrix elements as a function of the reduced time
$\theta\equiv H_0 \tau$, with the notation $ \dot{a}\equiv
da/d\theta$:
\begin{eqnarray}
\label{eqmot}
\rho_{0}^0 &=&-3 \ddot {a}/a,\\
\label{eqmot1}
 \rho_{i}^{i}&\equiv&\rho=-6\left( \ddot
{a}/a+\dot{a}^2/a^2\right).
\end{eqnarray}
The variation of Einstein's Lagrangian with an additional term $
\Delta L(R)\equiv f(R)$ gives
\begin{equation}
\label{einst} G_{i}^k=\frac{8\pi G}{H_{0}^2}T_{i}^{k}+\hat{T}_{i}^k;
\quad G_{i}^k=\rho_{i}^k-\frac{1}{2}\delta_{i}^k \rho.
\end{equation}
Here $T_{i}^k$ corresponds only to cold matter in the present
Universe and
\begin{eqnarray}
\label{T}
-\hat{T}_{i}^k &=& \left( \frac{\partial
f}{\partial \rho}\right) \rho_{i}^k - \frac{1}{2}\delta_{i}^{k}
f +\nonumber \\
&&{}\left( \delta_{i}^{k} g^{lm}-\delta_{i}^{l} g^{km}\right) \left(
\frac{\partial f}{\partial \rho}\right)_{;l;m}
\end{eqnarray}
specifies the effective quintessence with the nontrivial dependence
on curvature.

We proceed as in Ref.~\cite{Gur1} by introducing the new variable
\begin{equation}
\label{ya} y\equiv(\dot{a} a)^2
\end{equation}
which allows to reduce the order of the equations. Then the $i=k=0$
component of Eq. (\ref{einst}) leads to
\begin{eqnarray}
\label{00}
y+\left[ f_{,\rho}\left(
y-\frac{a}{2}\frac{dy}{da}\right) -\frac{a^4}{6}f(\rho)+ a y
\frac{df_{,\rho}}{da} \right]=\\ \frac{\rho_m}{\rho_{*}}
a^4\equiv\Omega_m a, \nonumber
\end{eqnarray}
where $\rho_m$ is the $a$-dependent cold dark matter (CDM) energy
density, $\rho_*=3 H_0^2/8\pi G$ is the critical density. Choosing
the value of the scale factor $a(\theta_0)$ equal to 1 at present,
one has $\rho_m=\rho_0/a^3$ and $(\rho_m/\rho_{*})a^4\equiv\Omega_m
a$.

In order to investigate the evolution of the cos\-molo\-gical model it
is enough to obtain the solution of Eq. (\ref{00}) with appropriate
initial conditions. But for interpretation of the solution and for
its comparison with observations it is necessary track for changes
of CDM energy density
 $\rho_m$ and quintessence energy density $\rho_v$
separately. For this purpose we use $i=k=0$ component of Einstein's
equations (\ref{einst})
\begin{equation}
G_0^0=\rho_0^0-\frac{1}{2}\rho=\frac{8\pi G}{H_0^2}\left(
\rho_m+\rho_v \right).
\end{equation}
Multiplying this equation by $a^4$ and using Eqs.~(\ref{eqmot}),(\ref{eqmot1}) and
(\ref{ya}), we have
\begin{equation}
y=(\rho_m+\rho_v) a^4/\rho_{*}
\end{equation}
accounting for the evolution of cold matter energy density. From
this we find for the quintessence energy density:
\begin{equation}
\frac{\rho_v}{\rho_{*}}=(y-\Omega_m a)/a^4.
\end{equation}
By solving Eq.~(\ref{00}), we
 can find the evolution of $\rho_v$. At
known $y$, the Hubble parameter
\begin{equation}
h(a)=\sqrt{y/a^4},
\end{equation}
the deceleration parameter of the Universe
\begin{equation}
\label{decel} q=-\frac{\ddot a a}{\dot a^2}=\frac{1}{2}\left(
1+\frac{3\delta_v w}{1+\delta_v} \right);
 \quad \delta_v=\rho_v/\rho_m,
\end{equation}
and the equation of state
\begin{equation}
\label{w} w=\left( \frac{2q-1}{3} \right) \left(
\frac{\delta_v+1}{\delta_v} \right).
\end{equation}

Here we will investigate  corrections to the
 Einstein-Hilbert   action of the form
\begin{equation}
\label{fr}
 f(R)=-\alpha \,R^n.
  \end{equation}

 In such  a case, Eq. (\ref{00}) can be
presented in the form
\begin{eqnarray}
\label{y}
  \beta\left[n(n-1)y''y+\frac{(1-n)}{2}(y')^2+n(4-3n)\frac{y'y}{a}
  \right]=\nonumber \\
  \frac{(y')^{2-n}}{a^{4-3n}}(y-\Omega_{m}\,a),\quad\\
  \beta=(-3)^{n-1}\alpha. \qquad \qquad \qquad \qquad \qquad \qquad \qquad \quad\nonumber
\end{eqnarray}
A general approach to investigation of the last equation is given
in~\cite{Gur1}.

From (\ref{y}) we see that a simple power in the
asymp\-totic solutions is absent for $n=2$ and $n=4/3$. These happen
to be the same parameters which were of special importance
 in a previous
theory of the early Universe. At $n=4/3$, one of us \cite{Gur2}
has obtained for the first time a
 cosmological model without
a singularity. That model passes through a regular minimum, has
inflationary stage and tends asymptotically to the classical
Fried\-mann solutions. In addition, the usefulness of an addi\-tional
term $R^2$ in the models of the early Universe has been pointed
out before. Therefore we analyze
 the possibility of using of such
powers for construction of models with variable parameters $q$ and
$w$.

\section{Models with $n=2$ and $n=4/3$}

These models have been considered in details in~\cite{fol}. The model with $n=2$
was often used in the theory of the early Universe.
 It describes the stage of fast
oscillations of $a(\theta)$ (the so called scalaron stage, which was
introduced by A. Starobinsky in \cite{Star2}). The damping of such
oscillations was connected with creation of unstable particles and
filling of the early Universe by a hot plasma.

Here we want to consider this model in the opposite regime when the
period of oscillations is commensurable with age of the Universe
$1/H_0$. One can easily see the oscillations of the model by
inserting the specified form of quintessence in (\ref{einst}) and
(\ref{T})
\begin{equation}
\frac{d^2\rho}{d\theta^2}+3\frac{\dot{a}}{a}\frac{d\rho}{d\theta}+(\rho+\Omega_m a)=0.
\end{equation}
The scalar curvature performs
 oscillations near the value of
$\rho$ which corresponds to the model of cold dust matter in the
Friedmann Universe.

The special feature of the model with $n=4/3$ consists in absence
of the scale factor $a$ in explicit form in the Einstein's
equations if matter is neglected \cite{Gur2}. This allows one to
find the general solution of Eq. (\ref{y}) with given $n$. One
can show that de Sitter's solution arises in the limit $a\gg 1$.
The curvature of this limiting solution is determined
 by the parameter
$\alpha$.

For a further analysis of these models and comparison  with
observable data see Section \ref{disc}

\section{The best-fit model}

In Eq. (\ref{y}) there are two parameters, $\beta$ and $n$,
deter\-mined by the observational data. These
parameters could be chosen according to different
requirements~\cite{Capoz,fol}. Here we will
choose parameter $n$ from requirement of  closeness of evolution
of our model to the classical solution for the  flat FRW
Universe with cold matter in the past \cite{Tok}. This fact  allows   this
scenario to be close to the scenario of the large scale structure
formation. This requirement can be realized at condition that the
classical Friedmann solution
\begin{equation}
\label{dust}
  y=\Omega_{m}\,a
\end{equation}
is a particular solution of Eq.(\ref{y}). It is easy to see
from Eq.(\ref{y}), the last condition is equivalent to the choice
of $n$ to be satisfying the equation
\begin{equation}
\label{eqn}
 n-1=2n(4-3n)  \qquad \qquad \qquad \qquad \qquad \qquad \qquad
\end{equation}
with roots
\begin{equation}
\label{roots}
n_1=1.295,\quad n_2=-0.129.
\end{equation}
The first of the roots leads to the type of models of
papers~\cite{Capoz,fol}, while the second root corresponds
the models with correction of the form $\propto \mu/R^{|n_2|}$
investigated in~\cite{Turner1}. As it will be shown
below, such a choice of $n$  approaches the model to the set of
the observational data in the best way.

\subsection{Behavior of the dust solution in
the $f(R)$-model in past}

After recombination, the evolution of the Universe has to be
described by the Friedmann model with the cold matter. The
 dynamics of expansion is determined by stability of the
dust solution in the model (\ref{y}). If the solution is stable,
then the model evolves in the way  very close to classical one.
However,  the observational data at $z<1$ does not correspond to
such a scenario, i.e. we are interested in  dust-like solutions
which  are not stable in the model (\ref{y}) but the perturbations
do not grow catastrophically fast. Otherwise, the model does not
provides a sufficiently long period with $q\simeq0.5$ at $z\gg1$
required for the large structure formation.

For investigation of behavior of the solution (\ref{dust}) we will
search  a perturbed solution in the form
\begin{equation}
\label{pert1} y=\Omega\, a(1+\psi),\quad \psi\ll1.
\end{equation}
 In this linear approximation and close to the recombination time
 ($a\ll1$) Eq. (\ref{y}) yields,
 \begin{eqnarray}
\label{psi0}
  n\,a^2\,\psi''+a\,\psi'(2n-0.5)=0,
\end{eqnarray}
with the damping solution $\psi=C_1+C_2\,a^{(\frac{1}{2n}-1)}$.
Hence the solution (\ref{dust}) for the $f(R)$-theory (\ref{fr}) with
$n=n_1$ satisfying Eq.(\ref{eqn}) asymptotically approaches to to
the flat FRW solution, i.e. it is stable and does not satisfies
the requirement stated above\footnote{\footnotesize It is
interesting to note that the mentioned exponents (\ref{eqn}) are
obtained in the recent paper~\cite{Multamaki} for the equation
equivalent to Eq.(\ref{y}). In the $f(R)$-theory, the given exponents
allow obtaining of  the solutions for cold dark matter coinciding
with ones in the classical Friedmann model of the Universe. Also,
it  has been shown in~\cite{Multamaki}  that these
solutions are stable within the framework of HOGT. Let us notice
 we slightly change exponent $n$ in the present work to obtain
weakly unstable solutions adequate to the observational data.
Authors are grateful to authors of paper~\cite{Multamaki} kindly
attracting our attention to their results.}.

As the next step on the path of  the choice of $n$,  we will look
for solutions of the $f(R)$-theories (\ref{fr}) with $n$ which is
a little different from $n_1$,
\begin{equation}
\label{dn1}
  n=n_1+\delta n,\qquad  \delta n\ll 1.
\end{equation}
It is efficient  to rewrite  the condition (\ref{eqn}) in the form
\begin{equation}
\label{eqnd} \delta\,(n-1)=2n(4-3n),\quad \delta=1+\epsilon,\quad
\epsilon\ll1.
\end{equation}
\begin{figure*} \centering
        \framebox[178mm]{\epsfxsize=155mm\epsfbox{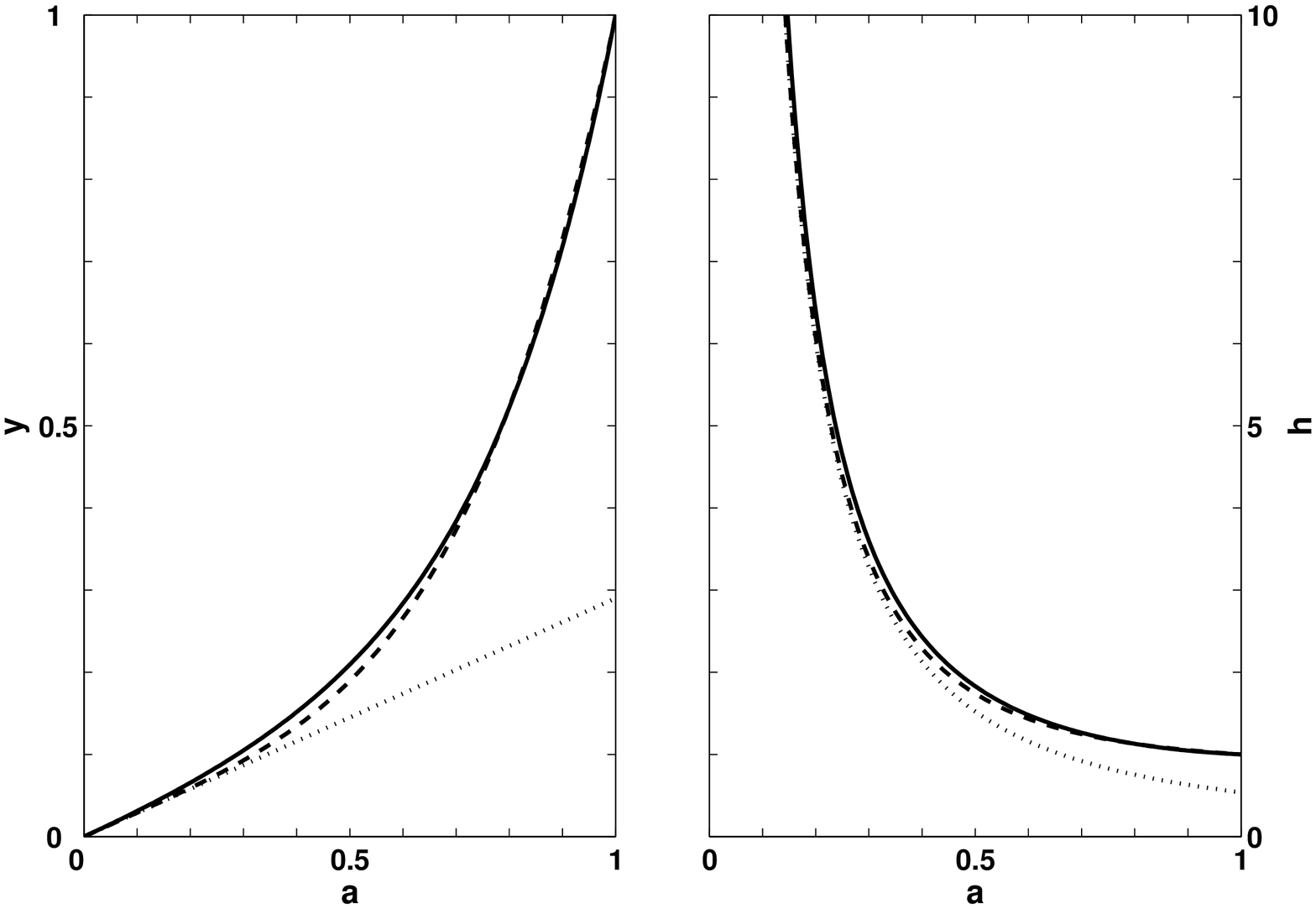}}
        \caption{\protect\small The evolution of  the variable $y$ (a left
panel) and the Hubble parameter $h$ (a right panel) with the scale
factor $a$  are represented by the solid line. The $\Lambda$CDM
solution and the cold matter solution are presented by dashed and
dotted line, respectively. Both graphs  correspond to the case
$\Omega_{m}=0.29.$}\medskip\hrule
\end{figure*}

 We will look for the perturbed dust-like solution in the form
(\ref{pert1}) near to recombination. Eq.(\ref{y}) for such a
case yields
\begin{eqnarray}
\label{psi1}
  n\,a^2\,\psi''+2a\,\psi'\left(n-0.5+\delta/4\right)
  +\left(\delta-1\right)=0.
\end{eqnarray}
Then after the change of variable  $\xi=\ln (a/a_{\ast})$ the Eq.
(\ref{psi1}) yields
\begin{equation}
 \label{psi1dot}
  n\ddot{\psi}+\left(n-1/2+\epsilon/2\right)
  \dot{\psi}+\epsilon=0,\,{\bf \dot{}}=d/d\xi.
\end{equation}

 The last equation has a solution
\begin{equation}
\label{pert}
    \psi=\left(C_1+\frac{2\,\epsilon}{1-2n}
    \ln\left(\frac{a}{a_{\ast}}\right)\right)+C_2\,a^{(\frac{1}{2n}-1)}.
\end{equation}
 The analysis of (\ref{pert}) has shown that the  requirement
 stated above is realized only for $\epsilon< 0$.

In this case, the modification of the exponent (\ref{eqnd}) is
determined by a small positive correction
\begin{equation}
\label{dn}
    \delta n \simeq -2\,\epsilon\, n_1(n_1-1)/(12n_1-1).
\end{equation}
 The numerical analysis has shown that behavior of solutions is
 sensitive to small changes of $\epsilon$ at $a\rightarrow1$. The
 last fact together  with a choice of parameter $\beta$ allows us
 to obtain a good enough  correspondence with the observational
data.

 As an illustration we  give  the results for the set of parameters
 $n=1.296,\; \beta=0.467$ fixed according to $\Omega_{m}=0.29, h_0=68$
  (the best fit to the CMB+SNe data presented in \cite{sahni1}).
  In a left panel of Fig.~1,  a solid line represents
  the evolution of
the variable $y$  with the scale factor $a$. At the beginning,
 it coincides with  the evolution of the dust model
  which is represented by a
 dotted line but further it deviates to the $\Lambda$CMD model
   represented by
 a dashed line. In a right panel,
one can see evolution of the Hubble parameter with the scale factor $a$ for  the mentioned
three models.

\subsection{The behavior of the solution in future}

The further expansion  of the Universe at $a\gg 1$ according to
Eq. (\ref{y}) leads to the negligible  effect of cold matter on the
solution behavior. In this case,   de Sitter
 solution $y=\Omega\, a^4$ is an asymptotic  solution
 of Eq.(\ref{y}). This solution corresponds to the
 constant Hubble parameter
 \begin{equation}
 \label{hOmega}
 h(a\rightarrow \infty)=\sqrt{\Omega}
 \end{equation}
  with  $\Omega$ defined from equality
\begin{equation}
\label{Omega}
 3\beta(2n-1)(n-1)/2=(4\Omega)^{(1-n)}/4
\end{equation}

  The inflationary solution
 is stable in the
 process of evolution of  the model. To show it,
 we shall look for a solution with perturbation in the form\\
 $y=\Omega a^4(1+\Phi),\quad \Phi \ll 1$. This ansatz yields
\begin{eqnarray}
\label{latepert}
 n_1a^2\Phi'' + [(n_1-1)/2+3n_1^2]a\Phi'+  \\
 6(2n_1^2-3n_1+1)\Phi=0. \nonumber
\end{eqnarray}
The change of variable  $a$ to variable $\xi$ (see Eq.(\ref{pert}))
yields
\begin{equation}
\label{latepert2} \ddot{\Phi}+A\dot{\Phi} +B\Phi=0,\quad
\dot{}=d/d\xi.
\end{equation}
where  coefficients  are $A=[3n_1-(1/n_1+1)/2]=2.99$,
$B=6(2n_1-3+1/n_1)=2.18$. This equation for perturbations have a
damped  solution indicating de Sitter solution  to be
 stable. The numerical analysis has shown that de Sitter solution is an
attractive solution.

As an example let us consider the evolution of the Hubble parameter  for the case
$\Omega_m=0.29$. In contrast to
 the Hubble parameter of $\Lambda$CDM
model  monotonically decreasing down to constant
$\sqrt{\Omega_{\Lambda}}$,  it reaches  a minimum $h_m \approx
0.978$ at $a\approx 1.15$ and after that increases up to the
asymptotic solution (\ref{hOmega}). It is interesting to note that  the
formula for the dimensionless Hubble parameter  $h(z)=h(a)$
  obtained from the obser\-vational data in \cite{sahni1} allows
   its extrapolation
to the  future $(a>1)$. At the parameters  mentioned in this paper,
the formula for $h(a)$ also predicts minimum of the Hubble
parameter at  $a\approx 1.45$ which is equal to $0.951$.
  The deceleration parameter $q$
 also passes a
minimum and  approaches to $-1$ with the growth of $a$. Therefore,
we live in a transitional epoch between the classical Friedmann
cosmology and a de Sitter cosmology.

\section{Discussion}
\label{disc}

  Hereafter we will present the comparison  of results of
   our model and   the observational data.

After fitting the model's parameters to the present observational
data - acceleration of the Universe, the Hubble parameter at
red-shift parameter $z=0$,
 and the age of the
Universe - there are no free parameters in the model. Its
predictions for large $z$ can be compared with observations.

In case of $n=2$,  the acceleration of the expansion
 changes
 at $z\approx 0.46$ to a
deceleration ($w\approx -0.72, \, q\approx 0$). Near $z=1$ the
variant of dust-like dark energy ($w=0$) is realized. It
corresponds to latest obser\-vati\-onal data~\cite{Sahni}.

A special feature of the model is that for $a\rightarrow 0$ the
variable  $y=(\dot a a)^2$   tends to a constant which is
equivalent to an evolution of the Universe filled by hot matter
($p=\epsilon/3$) in the Friedmann cosmology. When the solution is
oscillating, the inflation at $z\approx 0$ cannot be eternal
although the period of oscillations is comparable with the age of
the Universe.

In case of $n=4/3$, the crossover
from a
decelerated expansion
to an accelerated one takes place at $z\approx 0.30$ when
$w\approx -0.60$.

It is interesting to compare our model with the simplest $\Lambda$CDM model (for review see \cite{Sahni}).
Using
the notation
of section \ref{basic}, we  obtain
\begin{equation}
\label{lcdm_y}
y=\Omega_m \xi^{1/2}+\Omega_{v} \xi^2
\end{equation}
with the same values $\Omega_m\approx 0.3$ and $\Omega_v\approx 0.7$

Our results show the following: the age of the Uni\-verse is near $1/H_0\approx 13.7$ Gyr.
This age is larger than the age from the pure $\rho^2$-model ($\approx 11.4$ Gyr) and from the $\rho^{4/3}$-model ($\approx 10.8$ Gyr).
The parameter $w$ of
 the $\Lambda$CDM model calculating with use of (\ref{w})
has the value  $-1$. We may use
Eq.~(\ref{decel})
to determine the deceleration parameter
$q$  and compare it with the $\rho^2$ and $\rho^{4/3}$ models.
At $z=1 $, the deceleration parameter  is $q\approx 0.2$, whereas
 the $\rho^2$ and $\rho^{4/3}$ models have $q\approx 0.5$. This value is closer
to the observational data. At $z\approx 0.5$, the parameter $q$ in the $\rho^2$-model is close to zero and in the $\Lambda$CDM model
$q\approx -0.1$. Apparently,
the $\rho^2$-model corresponds better to observations.
The $\Lambda$CDM model is mathematically very simple.
But it leaves the value
of the $\Lambda$-term an unexplained fundamental constant.
 For dynamical
models of the $\Lambda$-term (for example, the $\rho^2$-model) this value evolves from a
large Planck value
in the early Universe to small value at present.
\begin{figure*} \centering
        \framebox[178mm]{\epsfxsize=145mm\epsfbox{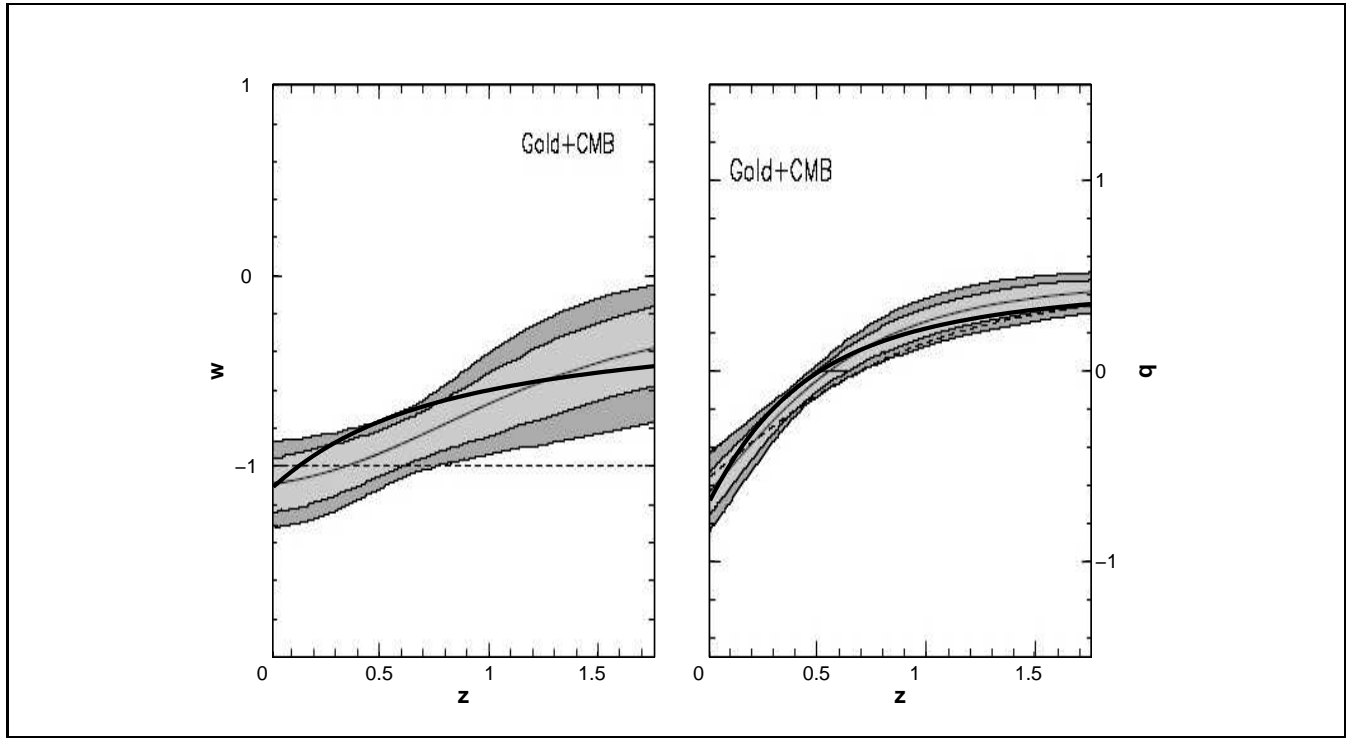}}
        \caption{\protect\small The comparison of results of $f(R)\propto
R^n$ model with $n=1.296$ and $\beta=0.467$ (the thick solid line)
to results of analysis of SNe+CMB data with $\Omega_{m}=0.29$ done
in~\cite{sahni1}. The evolution of the deceleration parameter with
redshift is shown  in a right panel and variation of equation of
state of DE is shown in a left panel. The best fit of SNe+CMB data
in such case is represented by the thin solid line, the $1\sigma$
and $2\sigma$ confidence levels are represented by the light and
dark grey contours, respectively, and $\Lambda$CDM is represented
by the dashed line.}
\medskip\hrule
\end{figure*}

   As a set of observational
   data,
   the analysis of SNe and CMB data from~\cite{sahni1}
    has been  used. In that paper authors have reconstructed
    the resent history of the Universe on the base of SNe  and CMB
     data in the model-independent way, only modeling DE
      by the hydro\-dynami\-cal equation of state
\begin{equation}
\label{w} p=w\rho, \quad w=(2q-1)/(3-\Omega_{m}/h^2).
\end{equation}
The cited paper presents two conceptions of the analysis of
the observational data: the first of them is the best fit to the data
which uses only the hydrodynamical describing of DE and does not
impose restrictions on the values of $\Omega_{m}$
 and $h_0$, while the second conception follows
 the priority of the  concordance
 $\Lambda$CDM model, so authors of~\cite{sahni1}
   put $\Omega_{m}=0.27\pm0.04$ and $h_0=0.71\pm0.06$.

 We  will give  the comparison of our results with both of them.
Also we notice that the analogy  of the ``hydrodynamical'' DE
(\ref{w}) is not so proper to the  higher order gravity theories,
hence one can expect the comparison over the  values $h$ and $q$
to be more informative than over the value $w$.

As it has been found in~\cite{sahni1}, the best fit values
are: $\Omega_{m}=0.385, h_0=60$. In the  model
     DE evolves in time
strongly enough.  For given $\Omega_{m}$ and $h_0$ we compared the
results of the $f(R)$-model with $ n= 1.2955 \,(\delta n=0.0002)$
and $\beta=0.273$ for the ``geometric equation of state''
parameter $w=(y-ay')/3(y-\Omega_{m}\,a)$ and the deceleration
parameter $q$ with the results of~\cite{sahni1}. In this
$f(R)$-model the age of the Universe is $14.9$ Hyr, the
deceleration parameter is $q_0=-0.91$ at present and  the
transition to acceleration occurs at $z=0.38$. Similarly to
results of~\cite{sahni1}, the $w_{DE} < -1$ at lower redshifts
($w_{DE0}= -1.53$), however , the evolution of equation of state
of ``geometric DE'' is more weak contrary to the results
of~\cite{sahni1}.

 However, if strong priors have been
imposed on $\Omega_{m}$ and $h_0$ (i.e. the $\Lambda$CDM model
priors: $\Omega_{m}=0.27\pm0.04$ and $h_0=0.71\pm0.06$), the
evolution of DE is extremely weak and in good agreement with the
$\Lambda$CDM model. The  best fit in the case is
$\Omega_{m}=0.29$ and there is a good enough coincidence of our
model and their analysis for parameters of the model $ n=
1.296\,(\delta
  n=0.001)$
and $\beta=0.467$ (see Fig.~2). The deceleration
parameter $q_0=-0.683$ at present, and the  deceleration was
changed by the acceleration at $z=0.51$ ($q_0=-0.63\pm0.12$ and
$z=0.57\pm0.07$ in \cite{sahni1}). The age of the
Universe in this case is 13.6 Hyr.

Thus, the $f(R)\propto \beta R^n$-model with  parameters $\beta$
and $n$ chosen according to the principles mentioned in
Introduction describes  the evolution of the Universe
 quite corresponding to the SNe+CMB data.

\end{document}